\documentclass[iop]{emulateapj}

\usepackage{natbib}
\usepackage{xcolor}
\usepackage{graphicx}
\usepackage{dcolumn}
\usepackage{bm}

\newcommand{\rgn}{($\gamma$,n)}
\newcommand{\rng}{($n$,$\gamma$)}

\newcommand{\rpg}{($p$,$\gamma$)}
\newcommand{\rpn}{($p$,$n$)}

\newcommand{\rdp}{($d$,$p$)}

\newcommand{\ovi}{$^{16}$O}
\newcommand{\ovii}{$^{17}$O}
\newcommand{\spro}{$s$-process}

\begin{document}

\title{Re-evaluation of the $^{16}$O($n$,$\gamma$)$^{17}$O cross section at astrophysical energies \\ and its role as neutron poison in the $s$ process}
\author{Peter Mohr}
\affil{Diakonie-Klinikum Schw\"abisch Hall, D-74523 Schw\"abisch Hall, Germany}
\affil{Institute for Nuclear Research (ATOMKI), H-4001 Debrecen, Hungary}
\email{WidmaierMohr@t-online.de; mohr@atomki.mta.hu}
\author{Christian Heinz}
\affil{II. Physikalisches Institut, Justus-Liebig-Universit\"at Giessen, Germany}
\author{Marco Pignatari}
\affil{E.A. Milne Centre for Astrophysics, Department of Physics \& Mathematics, University of Hull, HU6 7RX, United Kingdom}
\affil{Konkoly Thege Mikl\'{o}s Research Center for Astronomy and Earth Sciences, Hungarian Academy of Sciences, H-1051 Budapest, Hungary}
\affil{NuGrid collaboration, \url{http://www.nugridstars.org}}

\author{Iris Dillmann}
\affil{TRIUMF, Vancouver BC V6T 2A3, Canada}
\affil{Department of Physics and Astronomy, University of Victoria, Victoria BC V8W 2Y2, Canada}
\email{dillmann@triumf.ca}
\author{Alberto Mengoni}
\affil{ENEA Bologna, Italy}

\author{Franz K\"appeler}
\affil{Karlsruhe Institute of Technology (KIT), Campus Nord, Karlsruhe, Germany}


\begin{abstract}
The doubly-magic nucleus $^{16}$O has a small neutron capture cross section of just a few tens of microbarn in the astrophysical energy region. Despite of this, $^{16}$O plays an important role as neutron poison in the astrophysical slow neutron capture ($s$) process due to its high abundance. We present in this paper a re-evaluation of the available experimental data for $^{16}$O($n,\gamma$)$^{17}$O and derive a new recommendation for the Maxwellian-averaged cross sections (MACS) between $kT$= 5$-$100~keV. Our new recommendations are lower up to $kT$= 60~keV compared to the previously recommended values but up to 14\% higher at $kT$= 100~keV. 
We explore the impact of this different energy dependence on the weak $s$-process during core helium- ($kT$= 26~keV) and shell carbon burning ($kT$= 90~keV) in massive stars where \ovi~ is the most abundant isotope. 
\end{abstract}


\newpage

\section{Introduction}\label{sec:intro}
Theoretical models of the nucleosynthesis in stars can explain the origin of most nuclei beyond iron with a combination of processes involving neutron captures on short ("rapid neutron capture (r) process") or longer ("slow neutron capture (s) process") time scales \citep{bbfh57,Cam57,Cam57a,KGB11,thielemann:11}. These two processes contribute in about equal parts to the solar abundances beyond iron. A minor abundance fraction (corresponding to about 30 nuclei on the neutron-deficient side of the valley of stability between $^{74}$Se and $^{196}$Hg) is due to a superposition of several reaction mechanisms producing the so-called "$p$ isotopes" (see, e.g. \cite{RDD13}). 
The solar abundance peaks for all of these heavy element processes correspond to isotopes in the respective reactions paths with closed neutron shells ($N$ = 50, 82, and 126). However, the positions of these peaks are shifted due to the different regions of the reactions paths, and can be found for the $s$ process at $A=$90, 138, and 208.

The $s$-process distribution in the solar system can be divided into three components: a "weak" (60$<$$A$$<$90, \cite{PGH10}), a "main" (90$<$$A$$<$208, \cite{BTG14}), and a "strong" component (mostly including half of the solar $^{208}$Pb, \cite{GAB98}), corresponding to different astrophysical scenarios, temperatures, timescales, and neutron densities \citep{KGB11}. 

\subsection{The main and the strong $s$ process}\label{intro:mainstrong}
The main and strong $s$ process occur predominantly in low- and intermediate-mass (1$-$3~M$_\odot$) thermally pulsing asymptotic giant branch (TP-AGB) stars at different metallicities \citep{TGA04}. During the so-called Third Dredge-Up phase, protons from the convective hydrogen-rich envelope are mixed into the upper layers of the He intershell, which consists mostly of $^{4}$He, $^{12}$C and $^{16}$O. 
Here, the $^{12}$C can capture a proton to produce $^{13}$N, which $\beta^+$-decays to $^{13}$C. Depending on the amount of protons left, $^{14}$N might be produced via another proton capture on $^{13}$C. Thus, the ratio p/$^{12}$C will determine the amount of $^{13}$C and $^{14}$N \citep[][and references therein]{cristallo:09}. 
If a large amount of $^{14}$N is present, the  $^{14}$N$(n,p)$$^{14}$C reaction will capture all neutrons produced by the $^{13}$C($\alpha,n$)$^{16}$O reaction. In turn, if the amount of $^{13}$C is larger than the amount of $^{14}$N, neutrons will be available to be captured by $^{56}$Fe and heavier nuclei, and activate the $s$ process in the $^{13}$C pocket which is located in the radiative He-intershell between two convective thermal pulses. 
The following thermal pulse convectively mixes the $s$-process products into the He intershell and the temperature at the bottom of convective thermal pulses increases up to energies in the order of $kT$=25~keV. 
The reaction sequence $^{14}$N$(\alpha,\gamma)$$^{18}$F$(\beta^+)$$^{18}$O$(\alpha,\gamma)$ produces $^{22}$Ne, and the $^{22}$Ne($\alpha,n$)$^{25}$Mg neutron source can be partially activated, reaching peak neutron densities higher than  10$^{10}$~cm$^{-3}$. The following Third Dredge-Up event will enrich the stellar AGB envelope with fresh $s$-process material together with other light elements \citep[][]{herwig:05,karakas:14}.
%
Very recently new sensitivity studies for the main $s$-process and the relevance of the \ovi \rng \ovii~ reaction were presented by \cite{Kolo16} and \cite{Bist15}.

\subsection{The weak $s$-process}
The weak $s$-process component in the solar system is mostly made in massive stars (M$\gtrsim$10~M$_\odot$) during the convective core He- and shell C-burning phases and is responsible for the lighter $s$-process elements with A$<$90. In the earlier evolutionary stages of the He core, the $^{14}$N produced during the CNO cycle in the previous H burning stages is converted to $^{22}$Ne as described in Sec.~\ref{intro:mainstrong}. 
When the He concentration in the core drops below 10\% in mass fraction, the temperature rises up to $\sim$300~MK ($kT$= 25-30~keV) and efficiently activates the $^{22}$Ne($\alpha,n$)$^{25}$Mg reaction as main neutron source. 
The peak neutron density reaches a few 10$^{7}$~cm$^{-3}$ for $\approx$10$^4$ years. 
Since the $^{22}$Ne is not fully depleted during the He-burning stages, it can be re-activated in the following convective shell C-burning phase. The required $\alpha$-particles are produced via the $^{12}$C($^{12}$C,$\alpha$)$^{20}$Ne fusion reaction. At $\approx$1~GK ($kT$= 90~keV) the $^{22}$Ne($\alpha,n$)$^{25}$Mg reaction works efficiently and depletes most of the $^{22}$Ne in the timescale of few years. However, compared to the He core much higher peak neutron densities of up to a few 10$^{12}$~cm$^{-3}$ are reached, depending on the thermodynamic history of the C shell \citep[e.g.,][]{the:07,PGH10}. 

\subsection{Nuclear physics input}
While nowadays we know several details about the $s$-process production in stars, many stellar physics and nuclear physics uncertainties are affecting theoretical stellar model calculations. Concerning massive stars, robust $s$-process calculations need accurate neutron capture cross sections in the stellar energy range (E$_n$$\approx$0.1$-$500~keV), and a good knowledge of the $^{22}$Ne$+$$\alpha$ reaction rates. 

In the $s$-process path, a number of stable and also radioactive isotopes play key roles and require accurately measured neutron cross sections. In particular, theoretical prediction of $s$-process abundances depend on the production of $s$-only isotopes, which can be made only by the $s$ process, and on the production of unstable isotopes where the decay half-life is comparable with the neutron capture time scale (branching points of the $s$-process path, see e.g., \cite{KGB11,Bist15}). 


In addition, in the weak $s$-process isotopes with cross sections smaller than about 100$-$150~mbarn act as bottlenecks and induce a propagation effect on the reaction flow to heavier species \citep{PGH10}. 

Another crucial ingredient for reliable $s$-process calculations is an accurate knowledge about the neutron economy during the different phases. In the convective He-burning core about 70\% of all neutrons created by the $^{22}$Ne($\alpha,n$)$^{25}$Mg reaction are captured by light isotopes, while only the remaining 30\% are used to produce heavier $s$-process products beyond $^{56}$Fe. In C-burning conditions the amount of neutrons used for the $s$ process is $\lesssim$10\% \citep{PGH10}. 

Neutron poisons are defined as light isotopes that capture neutrons in competition with the $s$-process seed $^{56}$Fe. Well-known neutron poisons for the $s$ process in massive stars are $^{22}$Ne and $^{25}$Mg. 
In particular, uncertainties in the neutron-capture cross sections of neutron poisons are propagated to all the $s$-process isotopes beyond iron. The propagation effect due to light neutron poisons was first discussed by \cite{BG85} for $^{22}$Ne. For more recent discussions see \cite{PGH10}, and \cite{MKB12} for $^{25}$Mg and \cite{HPU14} for the Ne isotopes.


In general, the desired cross section uncertainty for all of these key isotopes at stellar temperatures is $<$5\%, which is up to now only achieved for a few isotopes between $A$=110$-$176 \citep{kadonis}. However, such small uncertainties can only be obtained under laboratory conditions, i.e. neglecting thermal excitations of the target under stellar conditions. This has to be taken into account by a theoretical correction of the laboratory result which may lead to an increased uncertainty of the stellar rate \citep{RMD11}.

In this paper we have re-evaluated the laboratory neutron capture cross section for a special neutron poison, $^{16}$O, in the energy range between $kT$= 5$-$100~keV. Sec.~\ref{sec:MACS} gives an overview about the existing data for the \ovi\rng\ovii~ reaction at stellar energies and explains the theoretical analysis. The new recommendation for the Maxwellian-averaged cross sections (MACS) between $kT$=5$-$100~keV can be found at the end of this Section. This cross section was implemented in the latest update of the neutron capture cross section database KADoNiS v1.0 \citep{kad10,kadonis}. Following this, we have investigated the influence of the new recommended values on the weak $s$-process nucleosynthesis. The production of \ovi~ in in rotating and non-rotating massive stars and its role as a neutron poison in the weak $s$-process is discussed in Sec.~\ref{sec:weak}. Finally, a summary and conclusions are given in Sec.~\ref{sec:summ}.

\section{Maxwellian-averaged cross section of the \ovi \rng \ovii\ reaction}\label{sec:MACS}
\begin{figure}[!htb]
\includegraphics[width=0.45\textwidth]{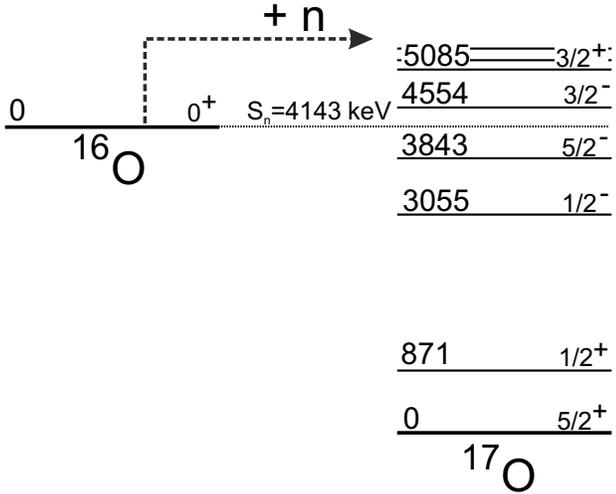}
\caption{\label{fig:level} Level scheme of \ovii\ up to an excitation energy $E\approx$ 6~MeV, based on the latest compilations \citep{TUNL}, and the \ovii ~neutron separation energy at $S_n$= 4143~keV. The relevant energy levels ($E_n$$<$ 1000~keV, corresponding to $E^*$$<$ 5143~keV) are labeled. All energies are given in keV.}
\end{figure}

The $Q$-value of the \ovi \rng \ovii\ reaction is relatively low ($Q = 4143$~keV) since \ovi\ is a doubly-magic nucleus. Consequently, the level density in the residual \ovii\ nucleus at the astrophysically relevant energy range is low, and the neutron capture cross section of \ovi\ is dominated by direct capture (DC). Below neutron energies of about 1~MeV, only two resonances have been observed. The level scheme of \ovii\ is shown in Fig.~\ref{fig:level}.

The neutron capture cross section of \ovi\ is given by the sum over all contributing final states, i.e.\ the $5/2^+$ ground state of \ovii\ and the excited states at $E^\ast = 871$~keV ($1/2^+$), 3055~keV ($1/2^-$), and 3843~keV ($5/2^-$). The latter state requires capture to a bound $f$-wave with $L = 3$ which cannot be reached by the dominating E1 transitions from incoming $s$- and $p$-waves. In addition, because of its excitation energy $E^\ast$ close to the $Q$-value of the \ovi \rng \ovii\ reaction, capture to the $5/2^-$ state is further suppressed by the small transition energy $E_\gamma = E + Q - E^\ast$; here $E$ is the neutron energy in the center-of-mass (c.m.) system. (Note that all energies $E$ are given in the c.m.\ system throughout this paper except explicitly stated.) Consequently, transitions to the $5/2^-$ state at 3843~keV have never been observed in the \ovi \rng \ovii\ reaction, and this state is not taken into account in the following.

\subsection{Available experimental data}
\label{sec:exp}
Neutron capture at thermal energies ($kT = 25.3$~meV) proceeds via $s$-wave capture. Two independent experimental results are available \citep{Jur63,McD77} which are in agreement. The weighted average of both results has been adopted: $\sigma = 190 \pm 18$~$\mu$b with a dominating 82~\% branch to the $1/2^-$ state at 3055~keV and a smaller 18~\% branch to the $1/2^+$ state at 871~keV \citep{McD77}. Very recently, a slightly smaller thermal capture cross section of $\sigma = 170 \pm 3$\,$\mu$b was obtained by \cite{Fire16}.

In the astrophysically most relevant keV energy region several experiments have been performed at the pelletron accelerator at the Tokyo Institute of Technology using neutrons from the $^7$Li\rpn $^7$Be reaction in combination with a time-of-flight technique. In a first experiment properties of the $3/2^-$ resonance at $E = 411$~keV ($E^\ast = 4554$~keV) were investigated \citep{Iga92}. In a second experiment the DC cross section at lower energies was measured \citep{Iga95}. A final experiment has covered the complete energy range from very low energies up to about 500~keV;
unfortunately, only a part of these data has been published \citep{Ohs00}, and the full data set is only available as private communication \citep{Nag00}. From the analysis of the latest data set it was noticed \citep{Ohs00} that the earlier determined radiation widths $\Gamma_{\gamma,0} = 1.80$~eV and $\Gamma_{\gamma,1} = 1.85$~eV in \cite{Iga92} are too large because strong contributions from non-resonant DC were not taken into account in \cite{Iga92}. 

In addition to neutron capture data, indirect information is available from other experiments. Ground state radiation widths $\Gamma_{\gamma,0}$ were derived from the \ovii \rgn \ovi\ photodisintegration reaction \citep{Holt78}: $\Gamma_{\gamma,0} = 0.42$~eV is found for the first resonance ($3/2^-$, $E = 411$~keV), and the second resonance ($3/2^+$, $E = 942$~keV) has
$\Gamma_{\gamma,0} = 1.0$~eV. Unfortunately, no uncertainties are given for these results. For completeness it should be noted that the analysis by \cite{Holt78} includes resonant and non-resonant $p$-wave photodisintegration and its interference; thus, it is not surprising that the result by \cite{Holt78} for the $3/2^-$ resonance at 411~keV is much lower than the value given in \cite{Iga92} where the significant non-resonant contribution was not taken into account.

Besides the partial radiation widths $\Gamma_\gamma$, a further essential ingredient in the analysis of the $^{16}$O$(n,\gamma)$$^{17}$O cross section is the total width $\Gamma$ of the lowest resonances. For the first $3/2^-$ resonance the adopted value is $\Gamma = 40 \pm 5$~keV \citep{TUNL} which is taken from an early transfer \ovi \rdp \ovii\ experiment \citep{Bro57}. Later a value $\Gamma = 45$~keV is reported from a $R$-matrix analysis of available total neutron cross sections on \ovi\ \citep{John67}; for the lowest resonance this analysis is based on earlier data by \cite{Oka55}. Finally, $\Gamma = 60 \pm 15$~keV is derived from neutron capture data \citep{AM71}. We adopt $\Gamma = 42.5 \pm 5$~keV for the following description of the $3/2^-$ resonance. The adopted width of the second resonance ($3/2^+$, $E = 942$~keV) is $\Gamma = 96 \pm 5$~keV which is the combined value of $\Gamma = 95 \pm 5$~keV from a direct width determination in the \ovi \rdp \ovii\ reaction \citep{Bro57}, $\Gamma = 97 \pm 5$~keV from a DWBA analysis of the \ovi \rdp \ovii\ transfer cross section \citep{And79}, and $\Gamma = 96$~keV or 94~keV from total neutron cross section data \citep{John73a,John73b,Str57}. Very recently, these adopted widths have essentially been confirmed, and the uncertainties have been reduced by about a factor of two \citep{Faest15}.

\subsection{Theoretical analysis}
\label{sec:theo}
The theoretical analysis is based on the direct capture (DC) model where the capture cross section is proportional to the square of the overlap integral of the scattering wave function $\chi(r)$, the bound state wave function $u(r)$, and the electromagnetic transition operator ${\cal{O}}^{E/M}$. Details on the DC formalism are given e.g.\ in \cite{Mohr93,Beer96}. The DC model considers the colliding projectile and target as inert nuclei which interact by an effective potential. As soon as this potential is fixed, the wave functions $\chi(r)$ and $u(r)$ can be calculated by solving the Schroedinger equation, and the overlap integral is well-defined.

The central potential is calculated from a folding procedure. An additional weak spin-orbit potential is taken in the usual Thomas form proportional to $1/r \times dV/dr$. The strengths of the central potential and the spin-orbit potential are adjusted to the energies of neutron single-particle states in \ovii\ by strength parameters
$\lambda$ for the central and $\lambda_{\rm{s.o.}}$ for the spin-orbit potential. In particular, this means $\lambda = 1.141$ for the $s$-wave potential from the adjustment to the $1/2^+$ state at $E^\ast = 871$~keV, and $\lambda = 1.117$ for the $d$-wave from the adjustment of the centroid of the $5/2^+$ ground state and the $3/2^+$ state at $E^\ast = 5085$~keV. For the $p$-wave the average value of $\lambda = 1.129$ was used. The spin-orbit strength $\lambda_{\rm{s.o.}}$ was adjusted to the splitting of the $5/2^+$ and $3/2^+$ states. By this choice of the potential, the $3/2^+$ state at
$E^\ast = 5085$~keV automatically appears as a resonance at $E = 942$~keV because it is included in the model space. A minor overestimation of the total width of this resonance ($\Gamma_{\rm{calc}} = 106$~keV, compared to $\Gamma_{\rm{exp}} = 96 \pm 5$~keV) does not affect the final MACS because the $3/2^+$ resonance does practically not contribute to the MACS at typical temperatures of the \spro . The ratio $\Gamma_{\rm{exp}}/\Gamma_{\rm{calc}} \approx 0.9$ close to unity clearly confirms the dominating single-particle structure of the $3/2^+$ resonance.

Usually, the calculated cross section in the DC model is finally scaled by the spectroscopic factor $C^2~S$ of the final state to reproduce the experimental capture cross section. Similar to a recent study of the mirror reaction \ovi \rpg $^{17}$F \citep{Ili08,Mohr12}, the present work uses the spectroscopic factor $C^2~S$ as an adjustable fitting parameter to adjust the theoretical
cross sections to the experimental capture cross sections. It has to be pointed out that a spectroscopic factor, which is determined in the above way, depends on the chosen potential \citep{Muk08}. However, the cross sections which result from the above fitting procedure are practically insensitive to the choice of the potential because the energy dependence of the neutron capture cross section is essentially defined by the available phase space, leading to cross sections proportional to $1/v$ ($v$, $v^3$) for $s$-wave ($p$-wave, $d$-wave) capture.

Because of the minor dependence of the calculated cross sections on the
underlying potential, earlier calculations have also reproduced
the experimental data with only small deviations, see e.g.\ 
\cite{Lik97,Men00,Kit00,Duf05,Yam09,Xu12,Dub13,Zhang15}.
However, no attempt was made
in these studies to adjust the theoretical cross sections to experimental data
for a $\chi^2$-based optimum description of the data. This is the prerequisite
for the determination of an experimentally based MACS. In some of the above cited
studies the comparison to experiment was restricted to the early data of
\cite{Iga95} which does not allow for a reliable $\chi^2$ adjustment in a
broader energy range.

As the $3/2^-$ resonance at $E = 411$~keV is not included in the model space of the DC model, the cross section of this resonance is added as a Breit-Wigner resonance. The total width is adopted as explained above ($\Gamma = 42.5$~keV), and the partial radiation widths $\Gamma_{\gamma,0}$ and $\Gamma_{\gamma,1}$ are adjusted to experimental neutron capture data. Because of the finite width, an additional interference term was also included in the analysis, similar to earlier work by \cite{Men00}.

The experimental data \citep{Nag00,Iga95} are average cross sections in a finite energy interval of about 20~keV (in the laboratory system). Therefore, the theoretical cross sections were averaged over corresponding intervals in the fitting procedure. Fortunately, it turned out that the derived spectroscopic factors are practically insensitive to that averaging, and even the derived radiation widths $\Gamma_{\gamma}$ of the 411~keV resonance remained stable within about 5~\%.

In detail, the following transitions in the \ovi \rng \ovii\ reaction were taken into account. Most of the transitions are electric dipole (E1) transitions. In few cases also M1 and/or E2 transitions have to be considered (details see below).
\begin{enumerate}
\item
$s$-wave capture to the $1/2^+$ state at $E^\ast = 871$~keV:\\
The contribution of this transition is well defined by the thermal (25.3~meV)
cross section of 34~$\mu$b \citep{McD77} and by the $1/v$ energy
dependence. At $kT = 30$~keV this transition is practically negligible
(0.03~$\mu$b).
\item
$s$-wave capture to the $1/2^-$ state at $E^\ast = 3055$~keV:\\
Similar to the previous transition, this cross section is well defined by the
thermal (25.3~meV) cross section of 156~$\mu$b \citep{McD77} and by the
$1/v$ energy dependence. At $kT = 30$~keV this transition remains very small
(0.14~$\mu$b).
\item
$p$-wave capture to the $5/2^+$ ground state:\\
A simultaneous fit of the spectroscopic factor of the ground state and the
ground state ratiation width of the $3/2^-$ resonance at 411~keV leads to
$C^2 S = 0.93 \pm 0.12$ and $\Gamma_{\gamma,0} = 0.30 \pm 0.07$~eV. The fit
is compared to the experimental data \citep{Nag00,Iga95} in
Fig.~\ref{fig:16O-MACS}. This transition contributes with about
10~$\mu$b to the MACS at $kT = 30$~keV.
\item
$p$-wave capture to the $1/2^+$ state at $E^\ast = 871$~keV:\\
A simultaneous fit of the spectroscopic factor of the $1/2^+$ state and the
partial radiation width of the $3/2^-$ resonance at 411~keV leads to
$C^2 S = 0.87 \pm 0.06$ and $\Gamma_{\gamma,1} = 0.53 \pm 0.05$~eV. The fit
is compared to the experimental data \citep{Nag00,Iga95} in
Fig.~\ref{fig:16O-MACS}. This is the dominating transition over the
entire astrophysically relevant temperature region. It contributes with about
25~$\mu$b at $kT = 30$~keV.
\item
$d$-wave capture to the $5/2^+$ ground state:\\
This transition includes the M1 resonance at $E = 942$~keV with
$\Gamma_{\gamma,0} = 1.0$~eV \citep{Holt78}. Despite the relatively large
width of 96~keV, the M1 cross sections remains negligible at lower energies;
e.g., at $kT = 30$~keV the M1 contribution is below 0.01~$\mu$b.
\item
$d$-wave capture to the $1/2^-$ state at $E^\ast = 3055$~keV:\\
This transition was not observed experimentally \citep{Nag00,Iga95}. The DC
calculation uses the small spectroscopic factor of $C^2 S = 0.012$ which is
derived from thermal $s$-wave capture to this state. The resulting contribution
remains negligible; e.g., at $kT = 30$~keV the MACS is below 0.01~$\mu$b.
\end{enumerate}

\subsection{Calculation of the MACS}
\label{sec:calc_MACS}
The total MACS of the \ovi \rng \ovii\ reaction is calculated from the contributions listed in the previous section. The MACS is given by \citep{BVW92} 
%
\begin{eqnarray}
<\sigma>_{kT} = \frac{2}{\sqrt{\pi}} \frac{1}{(kT)^2} ~ \times 
\quad \quad \quad \quad & \quad & \nonumber \\
\times \int_0^\infty \sigma(E) ~ E ~ \exp{[-E/(kT)]} ~ dE
\label{eq:MACS}
\end{eqnarray}
The cross section $\sigma(E)$ in the integrand of Eq.~(\ref{eq:MACS}) was calculated in small steps of 0.1~keV from 0.1 to 1000~keV. Then the Maxwellian-averaged cross section $<$$\sigma$$>$$_{kT}$ was calculated by numerical integration for thermal energies $kT$ between 1 and 150~keV; under these restrictions the exponential factor $\exp{[-E/(kT)]}$ is about $10^{-3}$ at the upper end of the integration interval which ensures sufficient numerical stability.

\begin{figure*}[!htb]
\begin{center}
\includegraphics[width=\textwidth]{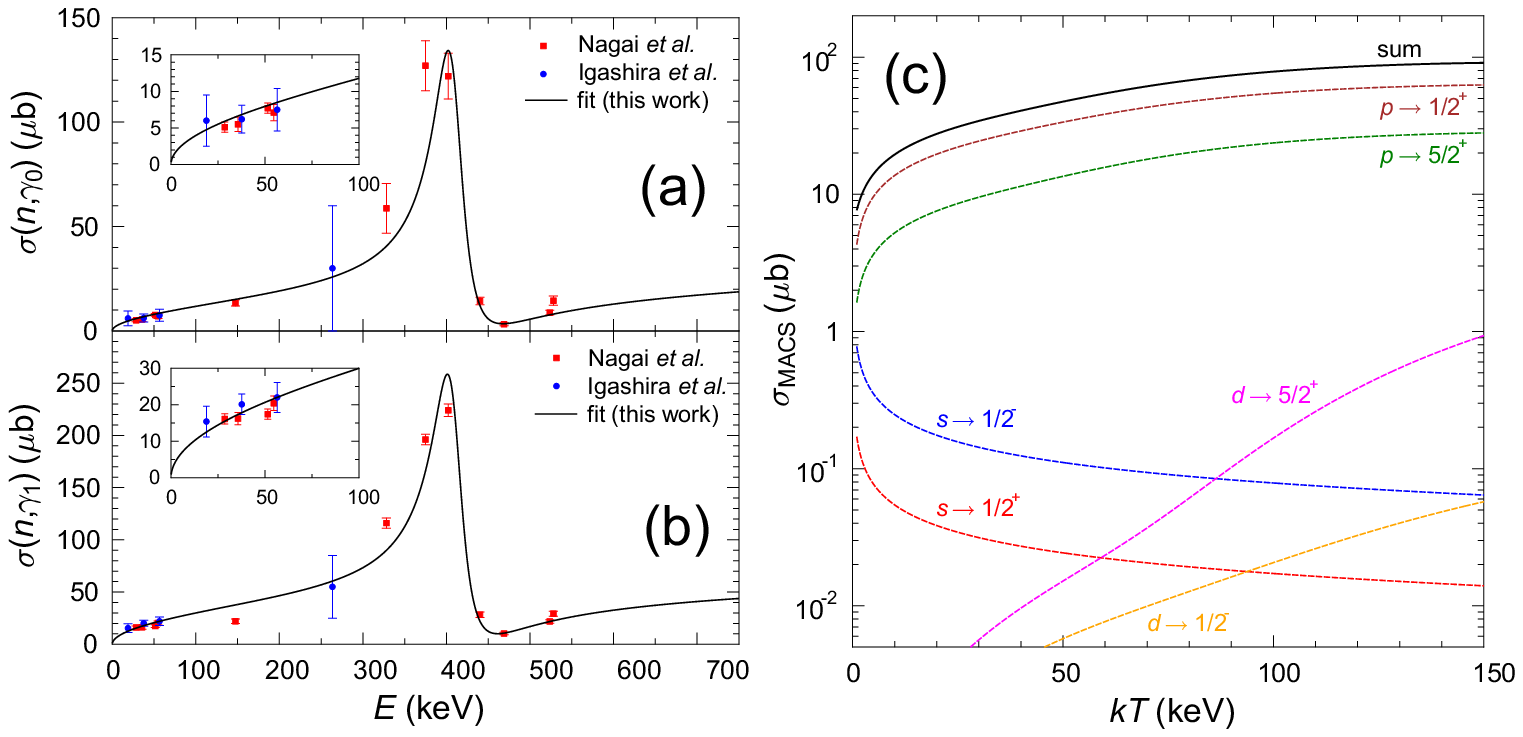}
\end{center}
\caption{\label{fig:16O-MACS}
(a): Experimental data for the ground state transition of the \ovi \rng \ovii\ reaction \citep{Nag00,Iga95} and the calculation with adjusted spectroscopic factor and ground state radiation width $\Gamma_{\gamma,0}$ of the $3/2^-$ resonance at 411~keV.  (b) Same as upper part, but for the transition to the first excited state ($1/2^+$, $E^\ast = 871$~keV). The insets enlarges the most relevant energy region below 100~keV.
(c) Maxwellian averaged cross section $<$$\sigma$$>$$_{kT}$ of the \ovi \rng \ovii\ reaction from the experimental data \citep{Nag00,Iga95} and the contributing transitions in logarithmic scale.
}
\end{figure*}

The MACS is calculated for all transitions listed in the previous section. The results are shown in Fig.~\ref{fig:16O-MACS}. It can be clearly seen that $p$-wave capture to the $5/2^+$ ground state and to the $1/2^+$ first excited state are the dominating contributions. Consequently, the uncertainty of the MACS is essentially defined by the uncertainties of the experimental data by \cite{Nag00,Iga95} which is slightly below 10~\% for the stronger transition to the $1/2^+$ state at 871~keV and slightly above 10~\% for the weaker ground state transition. The larger uncertainties for the radiation widths
$\Gamma_{\gamma,0}$ and $\Gamma_{\gamma,1}$ of the 411~keV resonance have only minor impact on the MACS at temperatures below $kT = 100$~keV. 

In total, this leads to an uncertainty of about 10~\% for the MACS over the
astrophysically relevant temperature range. However, the MACS and its
uncertainty are based essentially on one particular experiment which is not
fully published. An independent confirmation of the neutron capture data would
be very helpful. As a word of caution, it should be kept in mind that the only 
independent check to date is the ground state radiation width
$\Gamma_{\gamma,0}$ of the 411~keV resonance which was also measured by
\cite{Holt78} in the \ovii \rgn \ovi\  photodisintegration reaction. The
agreement ($0.30 \pm 0.07$~eV from neutron capture vs.\ 0.42~eV from
photodisintegration) is not perfect, and e.g.\ scaling the experimental data
by \cite{Nag00} to the average $\Gamma_{\gamma,0} = 0.36$~eV of the 411~keV
resonance would increase the MACS by 20~\%. In summary, we recommend the MACS
from the neutron capture data with an asymmetric uncertainty of $-10~\%$ from
the uncertainty of the neutron capture data and estimated $+20~\%$ from the
discrepancy for $\Gamma_{\gamma,0}$ from the two experimental
approaches. 

For completeness it should be pointed out that the stellar enhancement factor and the ground state contribution to the stellar MACS are very close to unity for the reaction under study because there are no low-lying excited states in the doubly-magic nucleus \ovi\ (see Fig.~\ref{fig:level}). Thus, an experimental determination of the stellar MACS is possible without additional theoretical uncertainties for the contributions of thermally excited states \citep{RMD11}.

\subsection{Renormalization with the new recommended $^{197}$Au$(n,\gamma)$$^{198}$Au cross section}
\label{sec:renorm}
$^{197}$Au is commonly used as reference for neutron capture cross section measurements. However, it is only considered a standard for thermal energies ($kT$= 25.3~meV) and in the energy range between 200~keV and 2.8~MeV \citep{Car09}. 
Recent time-of-flight measurements at n\_TOF \citep{Mas10,Led11} and at GELINA \citep{Mas14} revealed that the recommended $^{197}$Au$(n,\gamma)$$^{198}$Au cross section used in the previous KADoNiS databases was 5\% lower at $kT$= 30 keV than the new measurements. 

This previous recommendation was based on an activation measurement performed by the Karls\-ruhe group, which yielded a $<$$\sigma$$>$$_{kT}$= 582$\pm$9 mb at $kT$= 30 keV \citep{Rat88}. The extrapolation to higher and lower energies was done with the energy dependence measured at the ORELA facility \citep{Mac75}. The new TOF measurements \citep{Mas10,Led11,Mas14} are in perfect agreement with the recent ENDF/B-VII.1 evaluation \citep{endfb71} and with a new activation measurement by the group in Sevilla \citep{JiB14}.

The resulting new recommended dataset for $^{197}$Au in KADoNiS v1.0 \citep{kad10,kadonis} is given in Table~\ref{tab:xs}. For the astrophysical energy region between $kT$= 5 and 50 keV it was derived by the weighted average of the GELINA measurement and the n$\_$TOF measurement. The uncertainty in this energy range was taken from the GELINA measurement \citep{Mas14}. For the energies between $kT$= 60$-$100 keV the average of recent evaluated libraries (JEFF-3.2, JENDL-4.0, ENDF/B-VII.1) was used with the uncertainty from the standard deviation given in JEFF-3.2 and ENDF/B-VII.1. 

\begin{figure}[!htb]
\begin{center}
\includegraphics[width=0.45\textwidth]{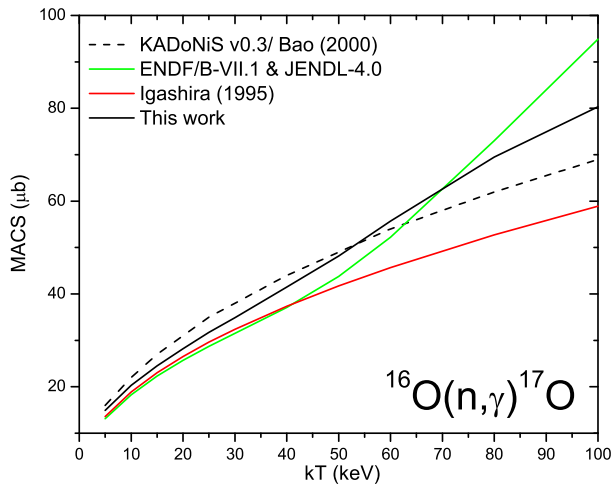}
\end{center}
\caption{\label{fig:MACS} Comparison of the MACS vs. kT energy dependence for $^{16}$O(n,$\gamma$)$^{17}$O from different sources.}
\end{figure}

All experimental cross section data for the $^{16}$O(n,$\gamma$)$^{17}$O reaction \citep{Iga92,Iga95,Nag00} have been obtained by normalization to the cross section of gold as given in ENDF/B-V.2 \citep{endf5,Mack67}. A comparison between the MACS for $^{197}$Au calculated with the ENDF/B-V.2 energy dependence and the new recommended MACS from KADoNiS v1.0 between $kT$= 5 and 100~keV shows differences between 1.7 and 5.9\% (Table~\ref{tab:xs}). This renormalization factor (ratio MACS KADoNiS v1.0/ ENDF/B-V.2) was applied in addition to our new MACS of the $^{16}$O(n,$\gamma$)$^{17}$O reaction which provides a very different energy dependence (see Fig.~\ref{fig:MACS}) compared to the previous recommendation. 

\begin{table*}[!htb]
\begin{tabular}{c|ccc|ccc}
 & \multicolumn{3}{c}{$^{197}$Au(n,$\gamma$)$^{198}$Au} & \multicolumn{3}{c}{$^{16}$O(n,$\gamma$)$^{17}$O}  \\
 \hline
$kT$ & $<$$\sigma$$>$$_{kT}$ (mb) & $<$$\sigma$$>$$_{kT}$ (mb) & Ratio & $<$$\sigma$$>$$_{kT}$ ($\mu$b) & $<$$\sigma$$>$$_{kT}$ ($\mu$b) & Ratio \\
(keV) & KADoNiS v1.0 & ENDF/B-V.2 &  KADoNiS/E-V.2 & This work & KADoNiS v0.3 &  This work/ v0.3 \\
\hline
5 & 2109 (20) & 1992 & 1.059 & 14.9 & 16 & 0.933 \\
8 & 1487 (13) & 1410 & 1.053 & 18.5 & - & - \\
10 & 1257 (10) & 1205 & 1.043 & 20.3 & 22 & 0.924 \\
15 & 944 (10) & 918 & 1.028 & 24.5 & 27 & 0.906 \\
20 & 782 (9) & 765 & 1.022 & 28.2 & 31 & 0.909 \\
25 & 683 (8) & 669 & 1.022 & 31.7 & 35 & 0.906 \\
30 & 613 (7) & 601  & 1.020 & 34.9 ($^{+7.0}_{-3.5}$) & 38 (4) & 0.920 \\
40 & 523 (6) & 512 & 1.021 & 41.5 & 44 & 0.944 \\
50 & 463 (5) & 455 & 1.017 & 48.2 & 49 & 0.983 \\
60 & 425 (5) & 415 & 1.024 & 55.7 & 54 & 1.031 \\
80 & 370 (4) & 361 & 1.025 & 69.5 & 62 & 1.121 \\
100 & 332 (4) & 324 & 1.025 & 80.3 & 69 & 1.164 \\
\hline
\hline
\end{tabular}
\caption{(Left column) New recommended MACS $<$$\sigma$$>$$_{kT}$ of $^{197}$Au(n,$\gamma$)$^{198}$Au from KADoNiS v1.0 \citep{kadonis} in comparison with the values calculated with ENDF/B-V.2 \citep{endf5}. 
(Right column) New recommended values for $^{16}$O(n,$\gamma$)$^{17}$O in comparison to the previous recommendations from KADoNiS v0.3 \citep{bao00,kad03}. The values given in brackets are the respective uncertainties.} \label{tab:xs}
\end{table*}

\subsection{Concluding remarks on the new MACS}
\label{sec:disc_theo}
The finally resulting $<$$\sigma$$>$$_{kT}$= 34.9$^{+7.0}_{-3.5}$~$\mu$b at $kT$= 30~keV for $^{16}$O$(n,\gamma)$$^{17}$O is slightly lower than the previously recommended value of 38~$\mu$b in the KADoNiS v0.3 database \citep{kad03} which is mainly based on a preliminary presentation by \cite{Nag95} using the experimental data of \cite{Iga95}. At very low temperatures the new MACS is close to the earlier result by \cite{Iga95} and about 7\% below the previous KADoNiS v0.3 recommendation \citep{bao00,kad03}. However, the MACS in Eq.~(4) of \cite{Iga95} neglects the resonance at 411~keV and the interference with the DC. Thus, it should not be used over a wider temperature range above $kT$$\approx$50~keV because it underestimates the MACS at $kT$$= $100~keV by up to 25\%.

The temperature dependence of the MACS in earlier work \citep{Iga95,bao00,kad03} follows approximately a $\sqrt{kT}$ dependence which results from the $v$ dependence of the $p$-wave capture cross section. Contrary to these earlier results, the recommended MACS now shows a somewhat stronger temperature dependence which is the consequence of the resonances at 411 and 942~keV and the significant constructive interference between direct and resonant capture below the 411~keV resonance. 
This leads to differences between the present MACS and the previous KADoNiS recommendation which goes from $\approx$$-$7\% at $kT$=5~keV to $\approx$$-$9\% at core He-burning temperatures up to $+$14\% at shell C-burning temperatures ($kT$=90~keV). We have carried out the weak $s$-process calculations in the following section with this new MACS and its revised energy dependence.

\subsection{Calculation of parameters for reaction libraries}
\label{sec:disc_7par}
Astrophysical reaction rate libraries (REACLIBs) consist traditionally of 8 different sections for different reactions. A detailed description of the REACLIB format can be found under http://download.nucastro.org/astro/reaclib or https://groups.nscl.msu.edu/jina/reaclib/db/. 

For historical reasons these libraries use a parametrization of seven parameters (a$_0$, a$_1$, a$_2$, a$_3$, a$_4$, a$_5$, and a$_6$) from which the reaction rate $N_A <\sigma v >$ (in cm$^3$/s/mole) for each temperature $T_9$ between 0.1 and 10~GK can be calculated via
\begin{eqnarray}
N_A < \sigma \, v > \, = \, \exp[a_0 + \frac{a_1}{T_9} + \frac{a_2}{T_9^{1/3}} + a_3
\cdot T_9^{1/3}  \nonumber \\ 
+ a_4 \cdot T_9 + a_5 \cdot T_9^{5/3} + a_6 \cdot \ln{(T_9)}]. \label{7para}
\end{eqnarray}

For our newly evaluated $^{16}$O$(n,\gamma)$$^{17}$O cross section up to $E_n$= 1~MeV we can provide this seven-parameter fit up to a temperature of $T_9$= 2, corresponding to $kT$= 173~keV. For temperatures of $T_9$= 2$-$10 a re-evaluation of higher-lying resonances is required which is beyond the scope of the present paper. For temperatures below $T_9 = 2$ the contribution of higher-lying resonances to the rate is less than 10\,\%. For temperatures $T_9 < 1$ the new rate is fully constrained by experimental data, and higher-lying resonances are completely negligible.

Table~\ref{tab:7par} compares the parameters from the Basel reaction rate library (http://download.nucastro.org/astro/reaclib)  with the fit parameters from the most recent JINA library (version 2.1, https://groups.nscl.msu.edu/jina/reaclib/db/). The entry in the Basel library provides a constant reaction rate (corresponding to a 1/$v$ energy-dependence) which was normalized to the previously recommended value from the KADoNiS database at $kT$= 30~keV of 38~$\mu$b (see Table~\ref{tab:xs}). The two entries in the JINA REACLIB provide a fit including the energy dependence of the previous recommendation in KADoNiS between $kT$=5 and 100~keV. 

\begin{table*}[!htb]
\begin{tabular}{c|ccccccc}
Ref. & a$_0$ & a$_1$ & a$_2$ & a$_3$ & a$_4$ & a$_5$ & a$_6$  \\
 \hline
this work & -1.355312E+01 &	5.460000E-02 &	-8.481710E+00 &	3.563536E+01 &	-4.033380E+00 &	2.048500E-01 &	-9.762790E+00 \\
Basel & 8.643563E+00 & 0 & 0 & 0  & 0 & 0 & 0 \\
JINA n1	& 3.388850E+00 & 0 & 0 & 0 & 0 & 0 & 0 \\
JINA n2	& 9.695150E+00 & 0 & 0 & 0 & 0 & 0 & 1.000000E+00 \\
\hline
\hline
\end{tabular}
\caption{Fit parameters for reaction rate libraries. Note that the parameters of the present work are only valid up to $T_9$= 2. Show are also the parameters from the Basel and the JINA REACLIB. The rate in the JINA REACLIB is divided into two non-resonant entries which have to be summed up.} \label{tab:7par}
\end{table*}

\begin{figure}[!htb]
\begin{center}
\includegraphics[width=1.0\columnwidth]{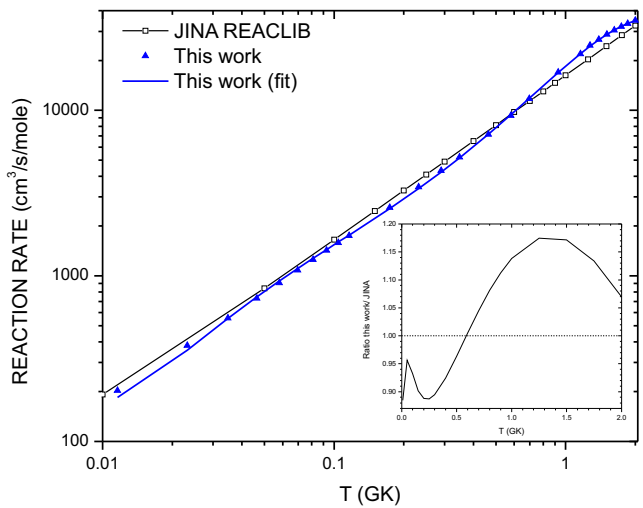}
\end{center}
\caption{\label{fig:rr} Comparison of the reaction rates up to T$_9$=2 calculated from this work and from the parameters in the JINA reaction library. The inset shows the ratio between this work and JINA.}
\end{figure}

Our fit was constrained between $T_9$=0 and 2. In Fig.~\ref{fig:rr} we compare our fit with the parameters from the JINA reaction rate library. The inset of the figure shows that the differences are $\pm$15\% in some regions. We emphasize again that our fit parameters are only valid up to a temperature of 2~GK and should not be used for any extrapolation beyond this temperature range. For completeness we also provide the tabulated values in the typical temperature grid of reaction rate libraries between $T$=0.1$-$2~GK in Table~\ref{tab:rr}.

\begin{table}[!htb]
\begin{tabular}{cc}
T (GK) & Rate (cm$^3$/s/mole)  \\
\hline
0.10 & 1542 \\
0.15 & 2223 \\
0.20 & 2910 \\
0.25 & 3628 \\
0.30 & 4385 \\
0.40 & 6033 \\
0.50 & 7852 \\
0.60 & 9820 \\
0.70 & 11908 \\
0.80 & 14079 \\
0.90 & 16296 \\
1.00 & 18520 \\
1.25 & 23879 \\
1.50 & 28573 \\
1.75 & 32262 \\
2.00 & 34771 \\
\hline
\hline
\end{tabular}
\caption{Tabulated reaction rate for the temperature range up to 2~GK.} \label{tab:rr}
\end{table}

\subsection{Comparison with evaluated libraries}
\begin{figure}[!htb]
\begin{center}
\includegraphics[width=1.0\columnwidth]{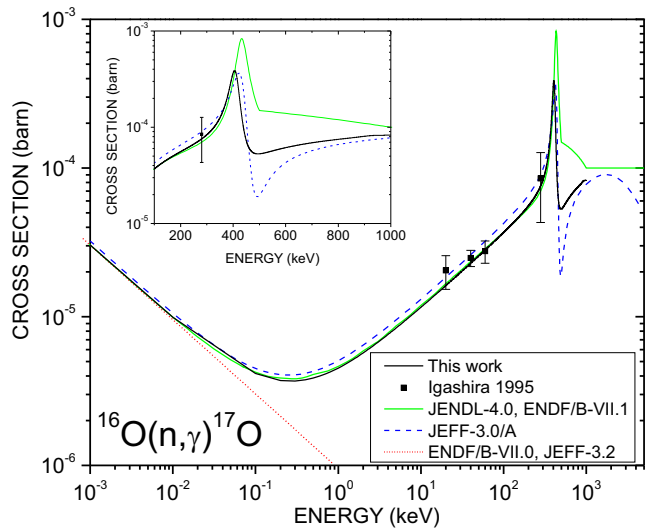}
\end{center}
\caption{\label{fig:eval} Comparison of the $^{16}$O(n,$\gamma$)$^{17}$O cross section from different evaluated libraries with the data from \cite{Iga95}.}
\end{figure}

In the following Table~\ref{tab:eval} we have compared our new result at $kT$= 30~keV with the values from recently evaluated databases and older compilations. The reason for the differences (and similarities) become clear when one looks at the respective capture cross sections in Fig.~\ref{fig:eval}.

The most striking difference comes from the fact that ENDF/B-VII.0 \citep{endfb7} (and earlier versions) as well as the latest JEFF-3.2 evaluation \citep{jeff32} use only a 1/$v$ extrapolation from thermal energies up to 20~MeV and thus strongly underestimate the MACS at stellar energies. The main reason is that the data from \cite{AM71} did not consider the DC contribution.

The JENDL-4.0 database \citep{jendl40} includes the data from \cite{Iga95} up to $E_n$= 280~keV, above this energy the statistical model code CASTHY \citep{casthy} is used but gives a "strange" (unphysical) shape for the region beyond the 411~keV resonance. The latest ENDF/B-VII.1 evaluation \citep{endfb71} adopts the JENDL-4.0 data.

The JEFF-3.0/A evaluation \citep{jeff30A} is lacking a more detailed information and cites a model calculation by A. Mengoni as private communication. From the shape of the data for this evaluation (see Fig.~\ref{fig:eval}), it can be deduced that the adopted cross section is as in \cite{Men00}.

\begin{table*}[!htb]
\renewcommand{\arraystretch}{1.2} 
\begin{tabular}{cccc}
Source & $<$$\sigma$$>$$_{30~keV}$ ($\mu$b) & Reference & Comments\\
 \hline
This work & 34.9 ($^{+7.0}_{-3.5}$) & & Re-evaluation\\
\hline
Igashira & 34 (4) & \cite{Iga95} & Experimental data\\
Allen \& Macklin & 0.2 (1) & \cite{AM71} & Experimental data\\
\hline
JEFF-3.2 & 0.17 & \cite{jeff32} & 1/$v$ extrapolation up to 20 MeV; \\ 
			 & & & up to 1~MeV data from \cite{JM64} \\
ENDF/B-VII.1 & 31.4 & \cite{endfb71}  & Uses JENDL-4.0 evaluation\\
JENDL-4.0 & 31.4 & \cite{jendl40} & \cite{Iga95}, \cite{casthy}\\
ENDF/B-VII.0 & 0.17 & \cite{endfb7} & 1/$v$ extrapolation up to 20 MeV; \\ 
			 & & & up to 1~MeV data from \cite{JM64} \\
JEFF-3.0/A & 35.8 & \cite{jeff30A} & Calculation by A. Mengoni \\
\hline
KADoNiS v0.0 & 38 (4) & \cite{bao00} & \cite{Nag95}, no 411 keV reson.\\
Beer et al. & 0.86 (10) & \cite{BVW92} & Reson. par. from \cite{MDH81}\\
\hline
\end{tabular}
\caption{Comparison of our MACS at $kT$= 30~keV with values from evaluated libraries and compilations, and the data from \cite{AM71} and \cite{Iga95}.\label{tab:eval}}
\end{table*}

\section{Impact on $s$-process simulations in massive stars}\label{sec:weak}

\subsection{The neutron absorber strength}
As mentioned in Sec.~\ref{sec:intro}, light isotopes capture a relevant fraction of the neutrons made by the $^{22}$Ne$(\alpha,n)$$^{25}$Mg reaction. This is due to the large abundance of some of these isotopes, which makes it more probable for them to capture a neutron despite the low neutron capture cross section. In this section we explore the $s$-process production and the impact of neutron poisons in trajectories extracted from the He core and from the C shell of a 25~M$_\odot$ stellar model with solar metallicity \citep{pignatari:13}. In the top panel of Fig.~\ref{fig:MF-NS} we show the respective time-evolution of the mass fractions of the most abundant species. 

We have added two "markers" to indicate when the $s$ process is activated in this plot: the hatched area during core He burning shows that the $^{22}$Ne$(\alpha,n)$ reaction is not yet activated until the temperature is high enough at about $t_{burn}$= 350000~y. At this point the production of $^{70}$Ge rises. During shell C burning, the temperature is already at the beginning high enough to run the $s$ process, and the $^{70}$Ge abundance is staying constant after a short rise. 

A qualitative measure for the strength of a neutron poison can be deduced by the integration of the abundance multiplied with the MACS at the respective burning energy $kT$. Ideally, this MACS includes all neutron-capture reaction channels, thus $(n,\gamma)$+$(n,p)$+$(n,\alpha)$. However, for the main neutron poisons in the He core and in  the C shell, the $(n,\gamma)$ cross section is always the most important neutron capture component. The following Table~\ref{tab:npoison} lists the values that have been used for the calculation of the neutron absorber strength in the bottom panel of Fig.~\ref{fig:MF-NS}. As can be seen, the $(n,p)$ and $(n,\alpha)$ channels are negligible, with exception of $^{14}$N$(n,p)$$^{14}$C.

If an isotope is classified as "neutron poison" or just as "neutron absorber" depends on if the captured neutron is recycled in a following reaction. If it is not recycled, then the neutron absorber is a neutron poison. Its high concentrations make \ovi~ one of the most efficient neutron absorbers when the $s$ process is activated in massive stars, both at the end of the convective He-core and in convective shell C-burning conditions \citep{PGH10}, despite its low neutron capture cross section. 

However, the $^{16}$O(n,$\gamma$)$^{17}$O reaction is followed by $\alpha$-capture on $^{17}$O via the two channels $^{17}$O$(\alpha,n)$$^{20}$Ne and $^{17}$O$(\alpha,\gamma)$$^{21}$Ne. The first channel is recycling neutrons captured by $^{16}$O back into the stellar environment, mitigating the impact of the $^{16}$O(n,$\gamma$)$^{17}$O reaction on the $s$-process neutron economy. Also the neutron capture channel $^{17}$O$(n,\alpha)$$^{14}$C  has a non-neglible contribution. 
Therefore, the relative efficiency between the two $\alpha$-capture channels on $^{17}$O is also crucial to define the relevance of $^{16}$O as a neutron poison. 
 
The importance of the $^{17}$O$(\alpha,n)$$^{20}$Ne and $^{17}$O$(\alpha,\gamma)$$^{21}$Ne rates was first discussed by \cite{baraffe:92} for the $s$ process in massive stars at low metallicity. A major source of uncertainty for theoretical nucleosynthesis calculations was given by the high uncertainty of the weakest channel $^{17}$O$(\alpha,\gamma)$$^{21}$Ne, with about a factor of a 1000 between the rates by \cite{caughlan:88} and \cite{descouvemont:93}. Recently, \cite{best:13} remeasured both $\alpha$-capture channels on $^{17}$O, providing more constraining experimental rates. 

It should be noted that $^{56}$Fe has the largest cross section compared to the light neutron poisons (absorbers) discussed here. However, its abundance at the beginning of each burning phase depends on the respective metallicity used in the simulations. For this reason we have not listed $^{56}$Fe in Table~\ref{tab:npoison} but show it for comparison in the plots for core He burning in Fig.~\ref{fig:MF-NS}.

\begin{table}[!htb]
\begin{tabular}{ccccc}
Isotope & Energy & \multicolumn{3}{c}{$<$$\sigma$$>$$_{kT}$ (mb)}\\
 & $kT$ (keV) & $(n,\gamma)$ & $(n,p)$ & $(n,\alpha)$ \\
 \hline
$^{12}$C & 25 & 0.0143 & - & - \\
		 & 90 & 0.0215 & - & - \\
\hline
$^{14}$N & 25 & 0.073 & 1.79 & negl. \\
		 & 90 & 0.043 & 5.30 & negl. \\
\hline
$^{16}$O & 25 & 0.0317 & - & - \\
		 & 90 & 0.0749 & - & - \\
\hline
$^{20}$Ne & 25 & 0.164 & - & negl. \\
		  & 90 & 0.518 & - & negl. \\
\hline
$^{22}$Ne & 25 & 0.053 & - & negl. \\
		  & 90 & 0.056 & - & negl. \\
\hline
$^{24}$Mg & 25 & 3.46 & - & - \\
		  & 90 & 2.61 & - & - \\
\hline
$^{25}$Mg & 25 & 5.21 & - & - \\
		  & 90 & 2.79 & - & - \\          
\end{tabular}
\caption{Maxwellian-averaged $(n,x)$ cross sections for the most abundant isotopes during core He and shell C burning. "negl." means that the calculated MACS is negligible compared to the other contributions. $^{4}$He is missing since all reactions lead to neutron-instable products, see text. The $(n,\gamma)$ MACS are taken from KADoNiS v1.0 \citep{kadonis}, the $(n,p)$ and $(n,\alpha)$ cross sections were taken from JEFF-3.0/A \citep{jeff30A}}. \label{tab:npoison}
\end{table}

\subsubsection{Core He burning}
At the beginning of core He burning the three most abundant species are $^{4}$He, $^{14}$N, and $^{20}$Ne. However, as indicated in Fig.~\ref{fig:MF-NS} by the hatched area, the temperature is not yet high enough to activate the $^{22}$Ne$(\alpha,n)$ reaction and thus the $s$-process is not started until $t_{burn}$$\approx$350000~y. At this point the $^{14}$N -- although the most important neutron poison in AGB stars due to the relatively large $^{14}$N$(n,p)$$^{14}$C cross section (see Table~\ref{tab:npoison}) -- has already been depleted and transformed into $^{22}$Ne by the reaction sequence $^{14}$N$(\alpha,\gamma)$$^{18}$F$(\beta^+)$$^{18}$O$(\alpha,\gamma)$$^{22}$Ne.

When the $s$-process is activated, the five most abundant isotopes are $^{16}$O, $^{4}$He, $^{20,22}$Ne, and $^{56}$Fe (since we used solar metallicity in our simulations).

The abundance of $^{16}$O is low at the beginning of the burning phase and originates mainly from previous star generations. However, it is copiously produced by the $^{12}$C$(\alpha,\gamma)$$^{16}$O reaction and becomes the most abundant isotope in the stellar core until O-burning conditions are reached in more advanced evolutionary stages. 

$^{4}$He is quickly depleted and can acts as neutron absorber but not poison. The reaction product $^{5}$He is a prompt neutron emitter and immediately recycles the captured neutron back into the system. The amount of $^{22}$Ne and $^{20}$Ne stays approximately the same during the $s$-process phase.

In the lower panel of Fig.~\ref{fig:MF-NS} the neutron absorber strength deduced from the abundance Y multiplied with the MACS at $kT$= 25~keV is shown. As discussed earlier, at the very beginning the $^{14}$N has the largest neutron absorber strength but it is very quickly depleted before the $s$ process is activated, and thus does not play a role as neutron poison during core He burning.

Once the $s$ process is started, $^{16}$O is the strongest neutron absorber, followed by $^{56}$Fe which is transformed into heavier $s$-process products. Towards the end of the burning phase, $^{25}$Mg produced by the $^{22}$Ne$(\alpha,n)$ reaction is the second strongest neutron poison.


\subsubsection{Shell C burning}
\begin{figure*}[!htb]
\begin{center}
\includegraphics[width=1.00\textwidth]{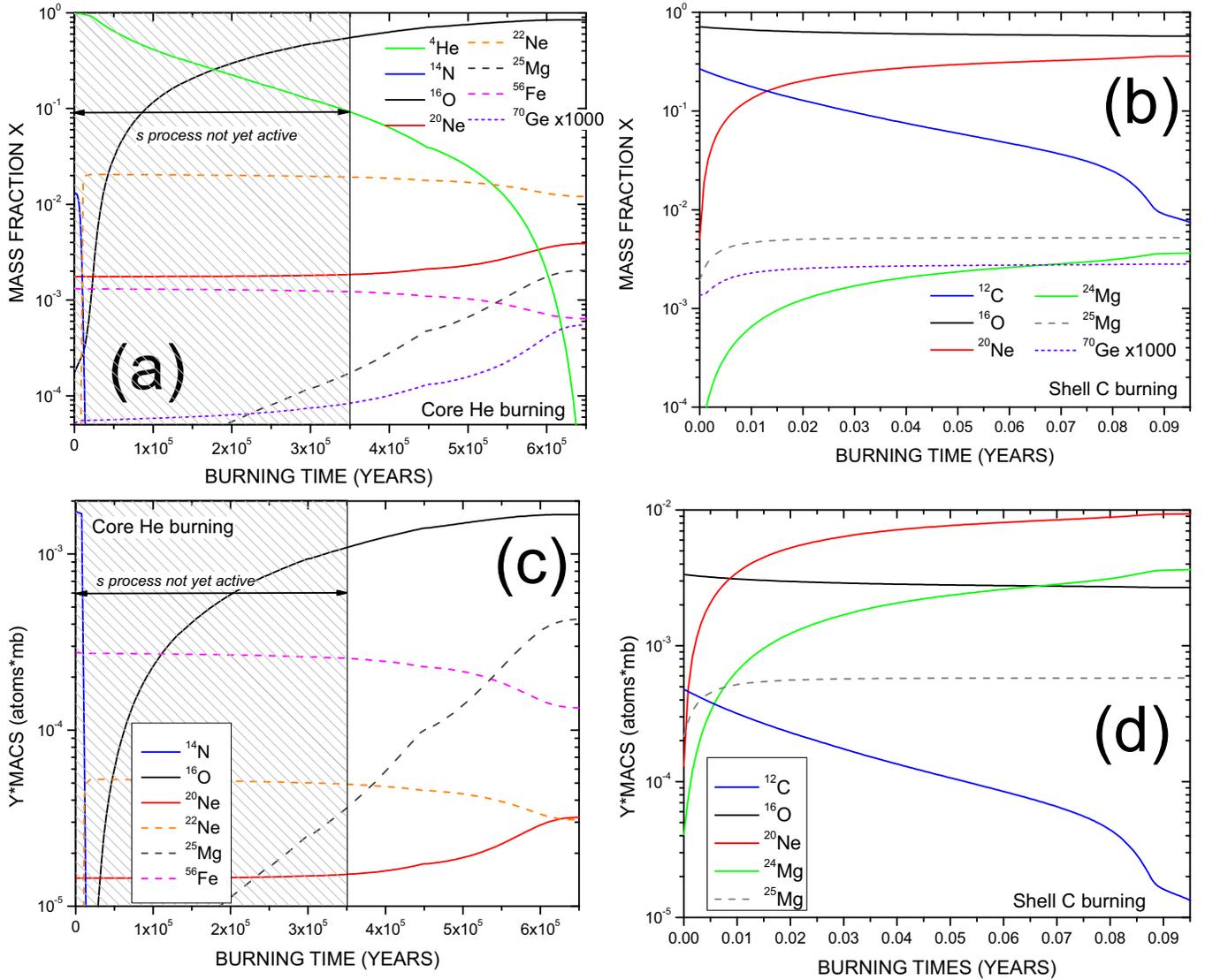}
\end{center}
\caption{\label{fig:MF-NS} Top: Time-evolution of the mass fraction X for $^{16}$O and other abundant isotopes during core He (left panel) and shell C burning (right panel). The mass fraction of $^{70}$Ge is plotted to indicate the start of the $s$ process. Bottom: Plot of the neutron absorber strength (abundance Y$\times$$<$$\sigma$$>$$_{25keV}$ or Y$\times$$<$$\sigma$$>$$_{90keV}$) for the aforementioned isotopes.}
\end{figure*}

At the beginning of shell C burning $^{16}$O and $^{12}$C are the most abundant species (see right panels in Fig.~\ref{fig:MF-NS}). $^{12}$C is quickly depleted by the reactions $^{12}$C($^{12}$C,$\alpha$)$^{20}$Ne, $^{12}$C($^{12}$C,$p$)$^{23}$Na, and to a smaller extent by $^{12}$C($\alpha,\gamma$)$^{16}$O and $^{12}$C($p,\gamma$)$^{13}$N.

Since the $^{12}$C($^{12}$C,$\alpha$)$^{20}$Ne reaction is the energetically most favourable, the abundance of $^{20}$Ne quickly rises. The other main fusion channel produces $^{23}$Na, which is efficiently depleted by the $^{23}$Na$(p,\alpha)$$^{20}$Ne reaction. The overall abundance of $^{16}$O does not change very much during this burning phase.

In the neutron absorber strength plot (lower panel of Fig.~\ref{fig:MF-NS}) one can see that $^{16}$O is the strongest neutron absorber only at the beginning of shell C burning and is quickly overtaken by $^{20}$Ne and later also by $^{24}$Mg.

\subsection{Weak $s$-process simulations}
We take one step further to simulate the effective weight of $^{16}$O as a neutron poison in $s$-process conditions in this work. Besides the recommendation of the stellar neutron capture reaction rate and its relative uncertainties we investigate the direct impact on weak $s$-process calculations in non-rotating and rotating massive stars.

\subsubsection{Non-rotating massive stars}
\begin{figure*}[!htb]
\begin{center}
\includegraphics[width=\textwidth]{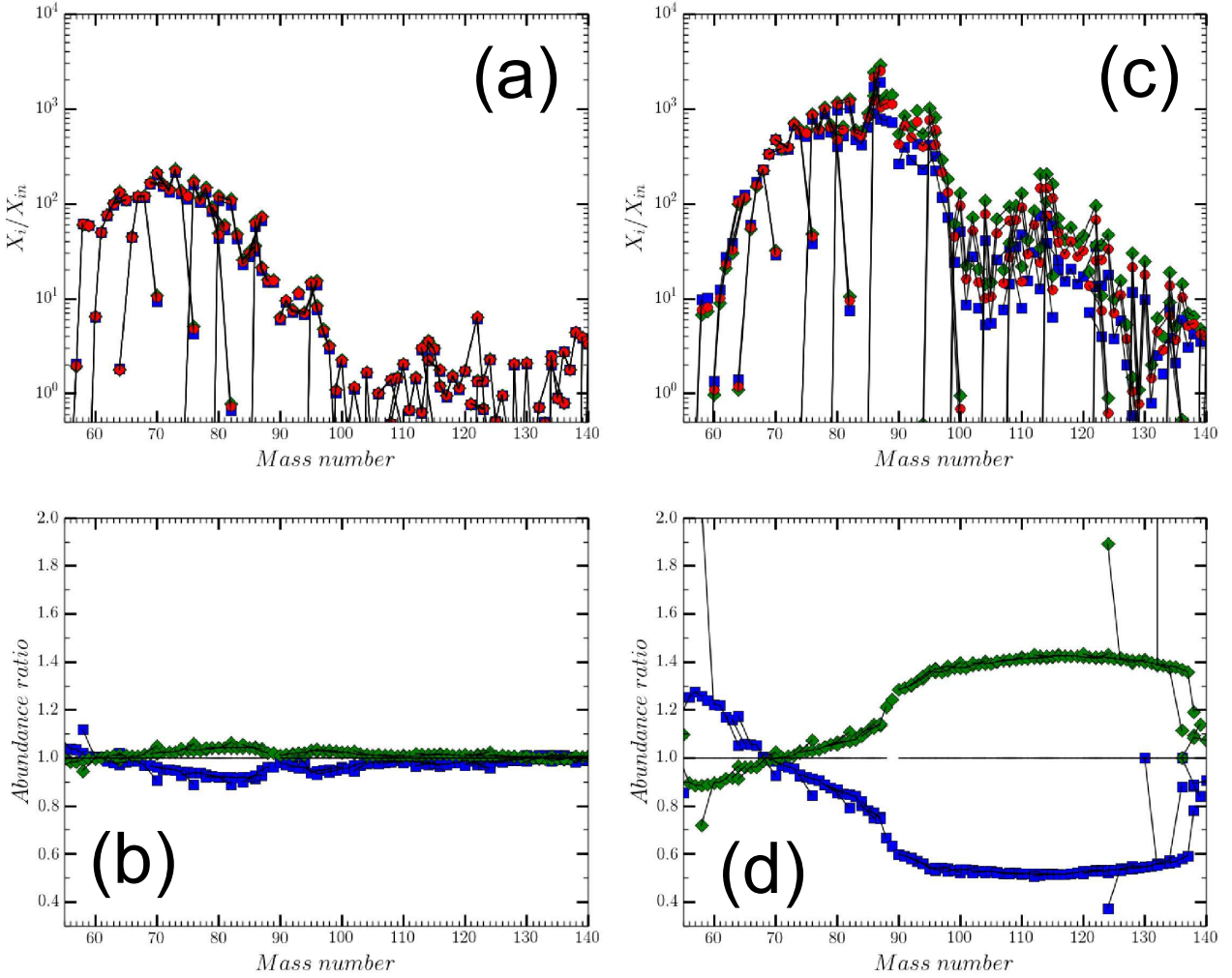}
\end{center}
\caption{\label{fig:prodfactors}
(a): Abundance distributions of the weak $s$ process in non-rotating massive stars calculated by using the same nuclear reaction network but varying the recommended $^{16}$O(n,$\gamma$)$^{17}$O MACS (red circles) within the upper limit (blue squares) and the lower limit (green diamonds).  (b): Ratio between the distributions obtained by using the upper limit and lower limit of the $^{16}$O(n,$\gamma$)$^{17}$O MACS and the recommended rate
(blue squares and green diamonds, respectively).
(c) and (d): As before but for the $s$-process in fast rotating massive stars at low metallicity.}
\end{figure*}

In the left panel of Fig.~\ref{fig:prodfactors} we show the effect of the $^{16}$O(n,$\gamma$)$^{17}$O uncertainties given in Table \ref{tab:xs} on the weak $s$-process distribution. Nucleosynthesis calculations were performed using the post-processing code PPN \citep{Pignatari:2012dw}. The single-zone $s$-process trajectory was exactracted from a complete 25~M$_{\odot}$ stellar model \citep{hirschi:08a}, calculated using the Geneva stellar evolution code GENEC \citep{eggenberger:08}. As expected, the obtained weak $s$-process distribution at solar metallicity is mostly efficient in the mass region 60$\lesssim$$A$$\lesssim$90, while its production is quickly decreasing beyond the neutron magic peak at $N$=50. The propagation effect of the $^{16}$O(n,$\gamma$)$^{17}$O MACS variation at the respective temperatures between its upper and lower limits is within 20\% over the $s$-process distribution.

\subsubsection{Fast rotating massive stars}
In the right panel of Fig.~\ref{fig:prodfactors} we explore the impact of the $^{16}$O(n,$\gamma$)$^{17}$O MACS in a trajectory representative for the $s$-process in fast rotating massive stars (e.g., \cite{pignatari:08,frischknecht:12}).
It is known that the $s$ process in massive stars is a secondary process, i.e. its efficiency decreases with the initial metallicity of the star. 
This is due to the fact that the neutron source $^{22}$Ne is made starting from the initial CNO nuclei, to the intrinsic secondary nature of the iron seeds, and due to the light neutron poisons \citep{prantzos:90, raiteri:92,pignatari:08a}.
On the other hand, \cite{pignatari:08} and \cite{frischknecht:12} showed that fast rotating massive stars at low metallicity can produce $s$-process yields which are orders of magnitude higher than in non-rotating stars. Recent galactical chemical evolution studies \citep{cescutti:13} have shown the potential relevance of this additional $s$-process source for the production of heavy elements in the early Galaxy. From a pure nucleosynthesis perspective, at low metallicity the main difference between the $s$ process in non-rotating massive stars and in fast rotators is due to the mixing of primary  $^{14}$N in the convective He core, which is rapidly converted to $^{22}$Ne via $\alpha$ capture \citep{meynet:06}. Therefore, in fast rotating massive stars the abundance of $^{22}$Ne is strongly enhanced, independently from the initial metallicity of the star.

For the calculations with fast rotators in the right panels of Fig.~\ref{fig:prodfactors} we use the same trajectories as for the "standard" weak $s$-process calculations 
but assume an initial metallicity of $Z$= 10$^{-5}$ (compared to the solar metallicity $Z_\odot$ taken for the non-rotating massive star simulations). 

Consistent with stellar calculations by \cite{hirschi:08b}, we use a concentration of primary $^{22}$Ne of 1\% in the convective He-burning core.
The $s$-process abundances at the Sr-Y-Zr neutron magic peak show the largest production factor. For $A \gtrsim 100$ the $s$-process production factors start to decrease. These results are consistent with previous calculations, adopting the new $^{17}$O$(\alpha,n)$$^{20}$Ne and $^{17}$O$(\alpha,\gamma)$$^{21}$Ne rates by \cite{best:13}.

The propagation of the $^{16}$O(n,$\gamma$)$^{17}$O uncertainties is much larger under fast rotator conditions, causing a variation up to about 40\% for $60 \lesssim A \lesssim 90$, and up to a factor of two for the $s$-process isotopes between the $N$=50 peak at $^{88}$Sr and the $N$=82 peak at $^{138}$Ba peak.
In particular, a higher (lower) $^{16}$O$(n,\gamma)$$^{17}$O MACS reduces (increases) the production of species heavier than A$\sim$90 and increases (decreases) the production of lighter heavy isotopes. 
This is due to the higher (lower) probability to capture neutrons by $^{16}$O, reducing (increasing) the $s$-process flow towards heavier species.
Compared to the weak $s$-process in non-rotating stars, the largest impact of $^{16}$O as a neutron poison is due to the fact that at low metallicity the typical secondary neutron poisons (e.g., $^{20}$Ne and $^{25}$Mg) are much weaker. Neutron poisons like $^{16}$O and the primary $^{22}$Ne itself become more relevant and therefore their uncertainties show a stronger propagation.

\section{Summary and Conclusions} \label{sec:summ}
We have re-evaluated the $^{16}$O(n,$\gamma$)$^{17}$O cross section at $kT$= 5--100~keV. Compared to the previously recommended MACS from \cite{bao00}, we derive a different energy dependence for $kT$$>$50~keV since the previous data neglected the contribution of a resonance at $E_n$= 411~keV and the interference with the DC component. This leads to an up to 16\% higher MACS at $kT$= 100~keV, close to the shell C-burning temperatures during the weak $s$-process.

An additional contribution to this change also comes from the recent re-evaluation of the $^{197}$Au(n,$\gamma$)$^{198}$Au cross section at astrophysical energies \citep{kadonis}. The previous $^{16}$O(n,$\gamma$)$^{17}$O cross section was measured relative to the Au cross section in the ENDF/B-V.II database \citep{endf5}, which is up to 5.9\% smaller at $kT$= 5~keV compared to the new recommended cross section given in this publication \citep{kadonis}.

Implementing this new recommended MACS of the $^{16}$O(n,$\gamma$)$^{17}$O reaction with its associated uncertainties into weak $s$-process simulations of fast-rotating massive stars, we observe a strong effect on the resulting abundance curve of up to 40\% for the mass region 60$\lesssim$$A$$\lesssim$ 90, and up to a factor of two for the $s$-process isotopes between the $N$=50 peak and the $N$=82 peak. This arises from fact that at lower metallicity the effect of secondary neutron poisons like $^{20}$Ne and $^{25}$Mg is much weaker, and the influence on the neutron economy is almost solely due to the change of the cross section of the neutron poison $^{16}$O.

This strong influence shows that a reduction of the experimental uncertainties in the production and destruction channels of neutron poisons is a crucial prerequisite for a better understanding of their role in the weak $s$-process, especially in fast rotating stars.

\acknowledgments
This work has been supported by the Hungarian OTKA (K101328 and K108459), the German Helmholtz Association via the Young Investigators projects VH-NG-627 and 327, and the Canadian NSERC Grants SAPIN-2014-00028 and RGPAS 462257-2014.
M.P. acknowledges significant support from NuGrid via NSF grants PHY 02-16783 and PHY 09-22648 (Joint Institute for Nuclear Astrophysics, JINA), NSF grant PHY-1430152 (JINA Center for the Evolution of the Elements) and
EU MIRG-CT-2006-046520. M.P. acknowledges support from the "Lendulet-2014" Programme of the Hungarian Academy of Sciences and from SNF (Switzerland), and from the UK BRIDGCE network.
NuGrid data is served by Canfar/CADC.
TRIUMF receives federal funding via a contribution agreement through the National Research Council of Canada.

\bibliography{refs-dillmann2}

\end{document}